\documentclass[table]{style/style}

\usepackage{microtype}
\usepackage{hyperref}
\usepackage{url}
\usepackage{booktabs}
\usepackage{graphicx}
\usepackage{lineno}
\usepackage{enumitem}
\usepackage{listings} %
\usepackage{svg}

\definecolor{ darkblue}{rgb}{0, 0, 0.5}
\hypersetup{colorlinks=true, citecolor=darkblue, linkcolor=darkblue, urlcolor=darkblue}

\usepackage{amssymb}
\usepackage{multirow}
\usepackage{bigdelim}
\usepackage{todonotes}
\usepackage{longtable}
\usepackage{tabularray}
\usepackage{wrapfig}
\usepackage[most]{tcolorbox} 
\usepackage{url}
\usepackage{xspace}
\usepackage{svg}
\usepackage[absolute]{textpos} %

\usepackage{fdsymbol}   %

\usepackage[utf8]{inputenc} %
\usepackage[T1]{fontenc}    %
\usepackage{url}            %
\usepackage{booktabs}       %
\usepackage{amsfonts}       %
\usepackage{nicefrac}       %
\usepackage{microtype}      %
\usepackage[table,dvipsnames]{xcolor}         %
\usepackage{amsmath}
\usepackage[most]{tcolorbox}
\usepackage{csquotes}
\usepackage{wrapfig}
\usepackage{siunitx}
\usepackage{graphicx}
\usepackage{arydshln}
\usepackage{wrapfig}
\usepackage{enumitem}
\usepackage{soul} %

\usepackage{multirow}
\usepackage{xspace}
\usepackage{adjustbox}
\usepackage{pifont}
\usepackage{caption}
\usepackage{makecell}
\usepackage{subcaption}
\usepackage{bold-extra}
\usepackage{url}
\usepackage{float}
\usepackage{pgf-pie}

\usepackage{hyperref}
\crefname{figure}{Figure}{Figures}
\crefname{section}{Section}{Sections}
\crefname{equation}{Equation}{Equations}
\crefname{appendix}{Appendix}{Appendice}
\crefname{table}{Table}{Tables}

\definecolor{linkcolor}{RGB}{0, 0, 128}
\hypersetup{
     colorlinks   = true,
     citecolor    = linkcolor,
     linkcolor    = linkcolor,
     urlcolor     = linkcolor,
}
\usepackage{pifont}%
\newcommand{\cmark}{\ding{51}}%
\newcommand{\xmark}{\ding{55}}%
\usepackage{listings}

\setlist[itemize]{leftmargin=*,itemsep=0em,parsep=0.3em,topsep=0.3em}

\DeclareUnicodeCharacter{2212}{\ensuremath{-}}

\usepackage{tikz}

\usepackage{setspace}

\usepackage{nicematrix}
\newcolumntype{L}[1]{>{\raggedright\let\newline\\\arraybackslash\hspace{0pt}}m{#1}}
\newcolumntype{C}[1]{>{\centering\let\newline\\\arraybackslash\hspace{0pt}}m{#1}}
\newcolumntype{R}[1]{>{\raggedleft\let\newline\\\arraybackslash\hspace{0pt}}m{#1}}
\newcolumntype{P}[1]{>{\centering\let\newline\\\arraybackslash\hspace{0pt}}m{#1}}
\colorlet{lightgray}{White!30!lightgray}
\colorlet{lightblue}{White!70!MidnightBlue}

\newcommand{\cls}{\texttt{[CLS]}}
\newcommand{\sep}{\texttt{[SEP]}}
\newcommand{\eos}{\texttt{[EOS]}}
\newcommand{\mamba}{\texttt{Mamba}\xspace}

\newcommand{\bert}{\texttt{BERT}\xspace}
\newcommand{\roberta}{\texttt{RoBERTa}\xspace}

\newcommand{\opt}{\texttt{OPT}\xspace}
\newcommand{\pythia}{\texttt{Pythia}\xspace}
\newcommand{\tfive}{\texttt{T5}\xspace}
\newcommand{\rwkv}{\texttt{RWKV}\xspace}
\newcommand{\matrixa}{\textbf{\emph{A}}}
\newcommand{\matrixb}{\textbf{\emph{B}}}
\newcommand{\matrixc}{\textbf{\emph{C}}}

\title{\LARGE RankMamba: Benchmarking \texttt{Mamba}'s Document Ranking Performance in the Era of Transformers}

\authorOne{Zhichao Xu}
\affiliation{University of Utah}
\contribution[]{Corresponds to: \texttt{zhichao.xu@utah.edu}}

\abstract{
Transformer structure has achieved great success in multiple applied machine learning communities, such as natural language processing (NLP), computer vision (CV) and information retrieval (IR). 
Transformer architecture's core mechanism\,---\,attention requires $O(n^2)$ time complexity in training and $O(n)$ time complexity in inference. 
Many works have been proposed to improve the attention mechanism's scalability, such as Flash Attention and Multi-query Attention.
A different line of work aims to design new mechanisms to replace attention. Recently, a notable model structure\,---\,\mamba, which is based on state space models, has achieved transformer-equivalent performance in multiple sequence modeling tasks. 
In this work, we examine \mamba's efficacy through the lens of a classical IR task\,---\,document ranking. 
A reranker model takes a query and a document as input, and predicts a scalar relevance score. 
This task demands the language model's ability to comprehend lengthy contextual inputs and to capture the interaction between query and document tokens. 
We find that \textbf{(1) Mamba models achieve competitive performance compared to transformer-based models with the same training recipe; (2) but also have a lower training throughput in comparison to efficient transformer implementations such as flash attention.}
We hope this study can serve as a starting point to explore \mamba models in other classical IR tasks. 
Our \href{https://github.com/zhichaoxu-shufe/RankMamba}{code implementation} is made public to facilitate reproducibility. Refer to~\cite{xu-etal-2025-state} for more comprehensive experiments and results, including passage ranking. 
}

\begin{document}

\maketitle

\section{Introduction}
Transformer structure has been the \emph{status quo} within NLP and IR community since the introduction of \texttt{Transformers}~\citep{vaswani2017attention} and \bert~\citep{devlin-etal-2019-bert}.
Compared to the classical RNN structure, the transformer structure is better at capturing long-range dependencies and also can be pre-trained on a large scale. 
However, in inference time, the transformer structure requires $O(n)$ time complexity and $O(n\cdot d)$ space complexity to store KV-cache, where $n$ denotes sequence length and $d$ denotes the size of the hidden state, making it less efficient compared to RNN. In addition, the transformer structure is also hard to extrapolate to context length longer than its pre-training length.

Recently, there has been a growing interest in developing alternative structures for modeling sequence data. For example, \rwkv~\citep{peng-etal-2023-rwkv} combines the efficient parallelizable training of transformers with the efficient inference of RNNs. Another notable structure is state space models (SSM)~\citep{gu2023mamba}, which are related to CNN, RNN and the classical state space models.
In essence, state space models compress the context into a smaller state, achieving $O(1)$ time complexity and $O(d)$ space complexity in inference time.
State space models are efficient in inference but also limited by the amount of information that can be compressed, i.e. the hidden state size. 
To mitigate this drawback, \citet{gu2023mamba} propose a novel selective state space model named \mamba.
\mamba\, addresses the computation problem of SSM with a selective scan method and corresponding hardware-aware optimization techniques, achieving performance close to transformer-based models of similar sizes while also demonstrating efficiency in training and inference.

Modern search systems typically consist of at least two stages: document retrieval and document ranking. During document retrieval, offline-built indexes are utilized to fetch a preliminary list of candidate documents. Subsequently, the document ranking task refines this initial retrieval list by modeling fine-grained query-document interactions.
Document ranking task requires the capability of language models (LM) to understand long context input and to capture the interaction between query and document tokens, at which transformer structure is good by construct, as the attention mechanism allows the query tokens to attend to document tokens. Given the success of \mamba in other sequence modeling tasks, we pose the following research question: 

\noindent
\begin{center}
\emph{RQ: Can state space models (e.g., \mamba) achieve comparable performance to transformer-based language models in the document ranking task?} 
\end{center}

\noindent
To comprehensively assess the performance of \mamba, we conduct a rigorous benchmarking study against transformer-based models characterized by varying structures, sizes, pre-training objectives, and attention patterns.
Correspondingly, we train neural reranker models following established training methodologies outlined in prior literature~\citep{boytsov2022understanding,gao2021rethink,ma2023fine}
This comparison allows us to evaluate the capabilities of different LMs in terms of performance and training efficiency. 
We find:
\begin{itemize}
    \item Encoder-only transformer-based LMs continue to demonstrate competitive performance in the document ranking task. For instance, \texttt{roberta-large} achieves 0.4434 MRR on MSMARCO Dev set, outperforming encoder-decoder LMs and decoder-only LMs of comparable size.
    \item \mamba models can achieve competitive performance, often matching or even surpassing transformer-based LMs of similar sizes. However, they also have lower training throughput compared to more efficient transformer implementations.
\end{itemize}

We first discuss the background of \mamba models in Section~\ref{sec:background}. 
We introduce our experiment setup in Section~\ref{sec:methodology} and report the experimental results in Section~\ref{sec:results}.
We review related works in Section~\ref{sec:related}, conclude this work and discuss limitations in Section~\ref{sec:conclusion}.

\section{Background on Selective State Space Models}
\label{sec:background}
\subsection{Structured State Space Sequence Models (S4)}
Prior works on state space models~\citep{gu2021combining,gupta2022diagonal,gu2022parameterization,smith2022simplified} are linear time invariance (LTI). 
We use Structured State Space Sequence Models (S4)~\cite{gu2021efficiently} as a representative example to illustrate the idea behind vanilla state space models.

State space models map a 1-dimensional function or sequence $x(t) \in \mathbb{R}$ to $y(t) \in \mathbb{R}$ through a latent state $h(t) \in \mathbb{R}^N$, where $t$ denotes a timestep. They are parameterized by $(\Delta, \matrixa, \matrixb, \matrixc)$ and define a \emph{continuous} sequence-to-sequence transformation as:
\begin{align}
    h'(t) &= \matrixa h(t) + \matrixb x(t)  & y(t) &= \matrixc h(t)
\end{align}
The above transformation can be \emph{discretized} as:
\begin{align}
    h_t &= \overline{\matrixa} h_{t-1} + \overline{\matrixb} x_t & y_t &= \matrixc h_t
    \label{eq:rnn}
\end{align}
The discretization of $\matrixa$ and $\overline{\matrixb}$ is defined by the \emph{discretization rule}, for example:
\begin{align}
    \overline{\matrixa} &= \exp(\Delta \matrixa) & \overline{\matrixb} &= (\Delta \matrixa)^{-1} (\exp(\Delta \matrixa)-I) \cdot \Delta \matrixb
\end{align}
% To make a concrete example, in natural language processing, $x_t$ and $y_t$ are parameterized by $D$-channel vectors, 
If we expand Equation~\ref{eq:rnn} with the whole sequence $x = (x_1, x_2, \ldots, x_n)$, we can arrive at:
\begin{align}
    \overline{\textbf{\emph{K}}} &= (\matrixc\overline{\matrixb}, C\overline{\matrixa\matrixb}, \ldots, C\overline{\matrixa}^{n-1} \overline{\matrixb})  & y &= x * \overline{\textbf{\emph{K}}}
    \label{eq:cnn}
\end{align}
Readers might have noticed that Equation~\ref{eq:rnn} looks like RNN while Equation~\ref{eq:cnn} looks like CNN, where $\overline{\textbf{\emph{K}}}$ is a large convolution kernel over the whole input sequence $x$. 
The parameterization of $(\Delta, \matrixa, \matrixb, \matrixc)$ is independent of input sequence $x$ and is fixed during all time steps, a property referred to as linear time invariance (LTI).
Structured state space models (S4) impose structure on the $\matrixa$ matrix to be computed efficiently.\footnote{An important note is that $\matrixa$ matrix also requires a specific initialization method: HiPPO~\cite{gu2020hippo} to memorize the context, the details of which will not be covered in this work.}
Existing works~\citep{gu2021efficiently,gupta2022diagonal,smith2022simplified} mostly deploy diagonal matrix, thus $\matrixa \in \mathbb{R}^{N \times N}$, $\matrixb \in \mathbb{R}^{N \times 1}$ and $\matrixc \in \mathbb{R}^{1 \times N}$ matrices are all represented by $N$ parameters.

Equation 1-4 can be generalized to $D$-channel features, i.e. $x_t, y_t \in \mathbb{R}^D$, a concrete example might be a word embedding of dimension $D$. In this case, computation of $\matrixa, \matrixb, \matrixc$ is applied to each channel independently. 

\subsection{Selective Scan Structured State Space Sequence Models (S6)}
State space models compress context into a smaller hidden state $h_t \in \mathbb{R}^N$, which limits the model's effectiveness.
\citet{gu2023mamba} propose to modify a subset of parameters $(\Delta, \matrixb, \matrixc)$ as input-dependent. 
This modification changes the model from time-invariant to time-varying, and poses challenges to the model's computational efficiency, as the model now cannot be trained in CNN mode.
Specifically, denote the context length as $L$ and batch size as $B$, training in RNN mode yields a total FLOPs of $O(BLDN)$, which cannot be trivially parallelized as each step depends on the output of the previous step. 

To tackle the computation problem, \citet{gu2023mamba} propose a hardware-aware optimization algorithm named \emph{Selective Scan}. 
It mainly comprises the following critical improvements:
\begin{itemize}
    \item Instead of computing $(\overline{\matrixa}, \overline{\matrixb})$ of size $(B,L,D,N)$ in GPU HBM (High-Bandwidth Memory), now SSM parameters $(\Delta, \matrixa, \matrixb, \matrixc)$ are loaded in GPU fast SRAM (Static Random Access Memory) to speed up the computing, and output of size $(B,L,D)$ are written back to HBM. This optimization technique is similar to Flash Attention~\cite{dao2024flashattention}.
    \item To avoid the sequential recurrence and to perform the computation in parallel, a work-efficient parallel scan algorithm~\cite{blellochpre,martin2018parallelizing} is used to amortize the computation to multiple threads within GPU. 
    \item Instead of saving the intermediate states, they are recomputed in the backward pass when the inputs are loaded from HBM to SRAM, which vastly reduces the memory requirements.
\end{itemize}

\citet{gu2023mamba} parameterize the corresponding selective state space model into a \mamba block, in which the selective scan structured state space sequence (S6) model only serves as a replacement for \texttt{Attention} component, while the majority of computation is handled by FFN layers and corresponding activation layers.
Combining the aforementioned optimization techniques, the parameterized \mamba block can be computed efficiently and achieve 5× higher throughput on long sequence lengths compared to Transformers of similar sizes, as reported in~\cite{gu2023mamba}.

\section{Methodology and Experiment Setup}
\label{sec:methodology}
We have briefly discussed state space models and the design choices of \mamba model. In this section, we focus on evaluating \mamba's efficacy in the document ranking task. 
To this end, we aim to address the following research questions:
\begin{itemize}
    \item RQ1: Between \mamba and transformer-based models of similar sizes, can \mamba perform comparably in document ranking task?
    \item RQ2: Can \mamba achieve higher training throughput compared to state-of-the-art efficient attention implementations?
\end{itemize}
Given the diverse array of existing transformer-based language models, we mainly focus on three distinct types: encoder-only models, decoder-only models, and encoder-decoder models. 
We elaborate on our design choices in this section.

\subsection{Task Formulation}
We focus on the reranking task, where the input to the reranker model is a search query and a candidate document from a first-stage retriever, and the model returns a scalar score indicating the relevance between the query and the document. 

Denote the reranker model as $f$, input query as $q \in \mathcal{Q}$, input document as $d \in \mathcal{D}$ and predicted relevance score as $s$,
\begin{equation}
    f(q, d) \rightarrow s \in \mathbb{R}
\end{equation}
Depending on the backbone language models we use, the exact ways to format $(q,d)$ are different. For models pre-trained with bi-directional attention, we prepend a \cls token to the input. On top of the backbone model, a fully connected layer (FFN) is randomly initialized, which takes the representation of the \cls token as input and predicts the relevance score $s$.
The input format is
\begin{equation*}
    \text{input} = \texttt{[CLS]} \lbrace q \rbrace \texttt{[SEP]} \lbrace d \rbrace \texttt{[EOS]}
\end{equation*}
where \sep and \eos are special tokens indicating separation and end of sequence, respectively.

Encoder-decoder models, i.e. \tfive\, family models can be used for feature extraction such as encoder-only models, and can be used for text generation such as decoder-only models.
As the encoder part is pre-trained with bidirectional attention, we use the same input format as encoder-only models. Then the FFN layer takes the representation of the first decoded token and predicts the relevance score. This setup is the same as prior works on adapting \tfive for the document ranking task~\citep{zhuang2023rankt5,ni-etal-2022-sentence}.

For decoder-only models pre-trained with unidirectional attention (often referred to as autoregressive LM), we use the following input format
\begin{equation*}
    \text{input} = \text{document: } \lbrace d \rbrace \backslash n \backslash n \, \text{query: } \lbrace q \rbrace \eos
\end{equation*}
The intuition for putting query $q$ at the later part of the input is to allow the query tokens to attend to prior document tokens to better capture relevance. 
On top of the backbone autoregressive LM, the randomly initialized FFN layer takes the representation of the \eos ~token as input and predicts the relevance score $s$.

For \mamba~models, as they are pre-trained with unidirectional information, we use the same input format as causal LMs.

\begin{table*}[t]
\vspace{0pt}
\centering
\caption{We list the compared backbone language models in terms of size, information, pre-training objective as well as model structure. 
NSP refers to the next sentence prediction objective.
$\mathbf{a}$ Note that \tfive\, family models are additionally multitask fine-tuned with GLUE tasks~\cite{wang-etal-2018-glue}. 
}
\resizebox{0.85\textwidth}{!}{
\begin{tabular}{
p{0.25\textwidth}
p{0.05\textwidth}
p{0.15\textwidth}
p{0.15\textwidth}
p{0.1\textwidth}
p{0.1\textwidth}
}
\toprule
\begin{tabular}[c]{@{}l@{}l} Model \\ \, \\ \end{tabular}  & \begin{tabular}[c]{@{}l@{}l} Size \\ \, \\ \end{tabular}  & \begin{tabular}[c]{@{}l@{}l} Information \\ \, \\ \end{tabular}  & \begin{tabular}[c]{@{}l@{}l} Pre-training\\ Objective \\ \end{tabular}  & \begin{tabular}[c]{@{}l@{}l} Model \\ Structure \\ \end{tabular}  & \begin{tabular}[c]{@{}l@{}l} Positional \\ Encoding \\ \end{tabular} \\ 
\rowcolor{lightgray}
\multicolumn{6}{l}{\emph{\textbf{Encoder-only Models}}} \\
% \texttt{distilbert-base-uncased}~\cite{sanh2019distilbert} & 66M & Bidirectional & MLM \& Distillation & Attention & Learned\\
\texttt{bert-base-uncased} & 110M & Bidirectional & MLM \& NSP & Attention & Learned \\
\texttt{bert-large-uncased} & 330M & Bidirectional & MLM \& NSP & Attention & Learned \\
\texttt{roberta-base} & 125M & Bidirectional & MLM & Attention & Learned \\
\texttt{roberta-large} & 355M & Bidirectional & MLM & Attention & Learned \\
\rowcolor{lightgray}
\multicolumn{6}{l}{\emph{\textbf{Decoder-only Models}}} \\

% \texttt{\texttt{EleutherAI/pythia-70m}~\cite{biderman2023pythia}} & 70M & Unidirectional & CLM & Attention & Rotary \\
\texttt{facebook/opt-125m} & 125M & Unidirectional & CLM & Attention & Learned \\
\texttt{facebook/opt-350m} & 350M & Unidirectional & CLM & Attention & Learned \\

\texttt{EleutherAI/pythia-160m} & 120M & Unidirectional & CLM & Attention & Rotary \\
\texttt{EleutherAI/pythia-410m} & 340M & Unidirectional & CLM & Attention & Rotary \\
\texttt{EleutherAI/pythia-1b} & 860M & Unidirectional & CLM & Attention & Rotary \\
\texttt{state-spaces/mamba-130m} & 130M & Unidirectional & CLM & \mamba & None \\
\texttt{state-spaces/mamba-370m} & 370M & Unidirectional & CLM & \mamba & None \\
\texttt{state-spaces/mamba-790m} & 790M & Unidirectional & CLM & \mamba & None \\
\rowcolor{lightgray}
\multicolumn{6}{l}{\emph{\textbf{Encoder-Decoder Models}}} \\
\texttt{T5-base} & 220M & Bidirectional & Denoising$^\mathbf{a}$ & Attention & Relative \\
\texttt{T5-large} & 770M & Bidirectional & Denoising & Attention & Relative \\
\bottomrule
\end{tabular}
}

\vspace{0pt}
\label{tab:models}
\end{table*}

\subsection{Training Objective}
Denote the relevant document to query $q_i$ as $d_i^+$ and sampled negative documents as $d_i^- \in \mathcal{D}_i^-$, the reranker is trained by optimizing a InfoNCE loss
\begin{equation}
    \mathcal{L} = \sum_{(q_i, d_i^+)} \frac{\exp (f(q_i, d_i^+))}{\exp (f(q_i, d_i^+)) + \sum_{j \in \mathcal{D}_i^-} \exp(f(q_i, d_j))}
\end{equation}
In practice, each mini-batch consists of multiple queries and corresponding documents, and loss is averaged over each query in the mini-batch before the optimizer takes an optimization step.

\subsection{Backbone Language Models}

\begin{table*}[t]
\vspace{0pt}
\centering
\caption{We list the training details and choice of hyperparameters. LR denotes learning rate, AMP denotes PyTorch Automatic Mixed Precision Training. Note that LoRA training requires a larger learning rate.}
\resizebox{0.75\textwidth}{!}{
\begin{tabular}{llllcc}
\toprule
Model & Size & LR  & AMP & Flash Attention & LoRA\\ 
\rowcolor{lightgray}
\multicolumn{6}{l}{\emph{\textbf{Encoder-only Models}}} \\
% \texttt{distilbert-base-uncased}~\cite{sanh2019distilbert} & 66M   & 2e-5 & FP16 & \xmark & \xmark\\
\texttt{bert-base-uncased} & 110M   & 2e-5 & FP16 & \xmark & \xmark\\
\texttt{bert-large-uncased} & 330M   & 1e-5 & FP16 & \xmark & \xmark\\
\texttt{roberta-base} & 125M   & 2e-5 & FP16 & \xmark & \xmark\\
\texttt{roberta-large} & 355M   & 1e-5 & FP16 & \xmark & \xmark\\
% \texttt{T5-base-Encoder} & 110M   & 2e-5 & BF16 & \xmark & \xmark \\
% \texttt{T5-large-Encoder} & 325M   & 1e-5 & BF16 & \xmark & \xmark \\
\rowcolor{lightgray}
\multicolumn{6}{l}{\emph{\textbf{Decoder-only Models}}} \\
% \texttt{\texttt{EleutherAI/pythia-70m}~\cite{biderman2023pythia}} & 70M & Unidirectional & CLM & Attention & Rotary \\
\texttt{facebook/opt-125m} & 125M   & 2e-5 & BF16 & \cmark & \xmark \\
\texttt{facebook/opt-350m} & 350M   & 1e-5 & BF16 & \cmark & \xmark \\

\texttt{EleutherAI/pythia-160m} & 120M   & 2e-5 & BF16 & \cmark & \xmark \\
\texttt{EleutherAI/pythia-410m} &  340M   & 1e-5 & BF16 & \cmark & \xmark \\
\texttt{EleutherAI/pythia-1b} &  860M   & 1e-4 & BF16 & \cmark & \cmark \\
\texttt{state-spaces/mamba-130m-hf} & 130M   & 2e-5 & BF16 & \xmark & \xmark \\
\texttt{state-spaces/mamba-370m-hf} & 370M   & 1e-5 & BF16 & \xmark & \xmark \\
\texttt{state-spaces/mamba-790m-hf} & 790M   & 1e-4 & BF16 & \xmark & \cmark \\
\rowcolor{lightgray}
\multicolumn{6}{l}{\emph{\textbf{Encoder-Decoder Models}}} \\
\texttt{T5-base} & 220M   & 2e-5 & BF16 & \xmark & \xmark \\
\texttt{T5-large} & 770M   & 1e-4 & BF16 & \xmark & \cmark \\
\bottomrule
\end{tabular}
}
\vspace{0pt}
\label{tab:hyperparameters}
\end{table*}

\begin{table*}[t]
\vspace{0pt}
\centering
\caption{We report official evaluation metrics for MSMARCO Dev, TREC DL19 and DL20. The highest number in each section is highlighted.}
\resizebox{0.8\textwidth}{!}{
\begin{tabular}{llrrr}
\toprule
\, & \, & MS MARCO Dev & TREC DL19 & TREC DL20 \\ 
\rowcolor{lightgray}
Model & Size & \multicolumn{1}{c}{MRR} & \multicolumn{2}{c}{NDCG@10} \\
\texttt{BM25(k1=4.46, b=0.82)} & - & 0.2767 & 0.5233 & 0.5061 \\
\hline
\texttt{bert-base-uncased} & 110M & \textbf{0.4126} & 0.6575 & 0.6147 \\
\texttt{roberta-base} & 125M & 0.4006 & 0.6591 & 0.6140 \\
\texttt{facebook/opt-125m} & 125M & 0.3878 & 0.6377 & 0.6176 \\
\texttt{EleutherAI/pythia-160m} & 120M & 0.3727 & 0.6490 & 0.6362 \\
\texttt{state-spaces/mamba-130m} & 130M & 0.4089 & \textbf{0.6652} & \textbf{0.6443} \\ 
\midrule
\texttt{T5-base} & 220M & 0.3670 & 0.6383 & 0.5932 \\ 
\midrule
\texttt{bert-large-uncased} & 330M & 0.4006 & 0.6591 & 0.6140 \\
\texttt{roberta-large} & 355M & \textbf{0.4334} & 0.6676 & \textbf{0.6423} \\
\texttt{facebook/opt-350m} & 350M & 0.3569 & 0.6431 & 0.6046 \\
\texttt{EleutherAI/pythia-410m} & 340M & 0.3750 & 0.6497 & 0.6169 \\
\texttt{state-spaces/mamba-370m} & 370M & 0.4250 & \textbf{0.6775} & 0.6394 \\ 
\midrule
\texttt{T5-large} & 770M & 0.3881 & 0.6473 & 0.6146 \\
\texttt{EleutherAI/pythia-1b} & 860M & 0.3921 & 0.6588 & 0.6386 \\
\texttt{state-spaces/mamba-790m} & 790M & \textbf{0.4201} & \textbf{0.6729} & \textbf{0.6489} \\

\bottomrule
\end{tabular}
}
\vspace{0pt}
\label{tab:results}
\end{table*}

We benchmark the performance of different models on three sizes: models with $\approx$ 110M parameters (e.g. \texttt{bert-base-uncased}); models with $\approx$ 330M parameters (e.g. \texttt{bert-large-uncased}) and models with $>$ 700M parameters (e.g. \texttt{T5-large} and \texttt{state-spaces/mamba-790m}).

For encoder-only models, we choose \bert~\citep{devlin-etal-2019-bert} and \roberta~\citep{liu2019roberta}; for decoder-only models, we include \pythia~\citep{biderman2023pythia}, \opt~\citep{zhang2022opt} and \mamba~\citep{gu2023mamba}, and we opt for \tfive~\citep{raffel2020exploring} as the representative encoder-decoder models.
The compared models are from different model families, varying in terms of pre-training objectives (e.g. causal language modeling, masked language modeling, denoising, next sentence prediction), information direction (bidirectional vs unidirectional), model structure (Attention vs \mamba) and Positional Encoding (no positional encoding, learned positional encoding, relative positional encoding and rotary positional encoding~\citep{su2024roformer}).
An important note is that we do not include instruction-finetuned language models such as \texttt{Flan-T5} models~\citep{chung2022scaling} for a fair comparison to other language model families. For a review of these models, refer to Table~\ref{tab:models}.

\subsection{Datasets}
We use MS MARCO document ranking dataset for training. The dataset consists of 3.2M documents sampled from search results of Bing search engine. We use the training queries and qrels from the official training set. 

Following prior practices~\citep{gao2021rethink,boytsov2022understanding,ma2023fine}, for each query and relevant document from the training set, we sample negative documents from the top-100 results returned by a sparse BM25 retriever implemented by Pyserini\footnote{\url{https://github.com/castorini/pyserini}}~\citep{lin2021pyserini}.
In practice, we fix the random seed and sample 7$\times$2 negative documents for each qrel in the official training set, leading to in total 734,008 training samples, each consisting of 1 relevant document and 7 hard negative documents. 
Notably, we use the setup recommended by \cite{gao2021rethink} to sample hard negatives from the top-100 results returned by the first-stage retriever (BM25 in our case).
The sample training set is used throughout all experiments to ensure fair comparison and reproducibility.

We use the official document ranking Dev set for evaluation. We also use TREC DL19~\citep{craswell2020overview} and DL20~\citep{craswell2021overview}.
The official Dev set consists of 5,193 queries, with 1 relevant judgment each, while DL19 and DL20 consist of 43 and 45 queries respectively with denser relevance judgments.

We use the reranking setup in evaluation as prior practices~\citep{boytsov2022understanding,gao2021rethink,ma2023fine} where a first-stage retriever BM25 retrieves top-100 documents and our trained reranker models perform the reranking. 
We report Mean Reciprocal Rank@100 for Dev set and NDCG@10 for DL19 and DL20 to be consistent with the official evaluation.
Deploying better first-stage retrievers can lead to better reranking performance but is beyond the scope of this study.

\subsection{Training Details and Hyperparameters}

We truncate the input to 512 tokens to fit in the context window of \bert and \roberta and \tfive models (also referred to as FirstP by~\citet{dai2019deeper}). For autoregressive LMs (\opt, \pythia and \mamba), although they have a longer context length, we also use 512 tokens to keep a fair comparison. 
We use the official implementation of \mamba models\footnote{\url{https://github.com/state-spaces/mamba}} and weights of other models are downloaded from Huggingface.\footnote{\url{https://huggingface.co/models}}

For all experiments, we optimize the reranker model with AdamW optimizer~\citep{loshchilov2018decoupled}, with betas=(0.9, 0.999), eps=1e-08, weight\_decay=0.01. 
In our preliminary experiments, we noticed models larger than 300M can suffer from overfitting, therefore we use a smaller learning rate for larger models. 
We use a linear warmup scheduling where we gradually increase the learning rate in the first 1000 training steps and decrease to zero until the end of training.
We use PyTorch native mixed precision training and Flash Attention 2~\cite{dao2024flashattention} for higher throughput.
For models larger than 700M, i.e. \texttt{T5-large}, \texttt{EleutherAI/pythia-1b} and \texttt{state-spaces/mamba-790m}, we use LoRA~\citep{hu2021lora} to reduce memory consumption and accelerate training.

We train the reranker models for 1 single epoch, as recommended by \cite{ma2023fine}. Through the course of training, for each reranker model, we save 10 intermediate checkpoints to monitor the training trajectory and report the performance of the final checkpoint.
Refer to Table~\ref{tab:hyperparameters} for a detailed listing of hyperparameters and training details.

\section{Result and Analysis}
\label{sec:results}
\noindent
\textbf{Document Ranking Performance.} 
We report the ranking performance in Table~\ref{tab:results}. 
We notice that encoder-only language models such as \bert and \roberta can achieve strong performance. For example, within models $\approx$ 110M and models $\approx$ 330M parameters, \texttt{bert-base-uncased} and \texttt{roberta-large} achieve the highest MRR on MSMARCO Dev dataset.

For decoder-only LMs such as \opt and \pythia, we notice an interesting discrepancy between performance on datasets with in-depth relevance judgments (DL19 and DL20) and MSMARCO Dev which has shallow judgments (1 relevant document per query). A similar trend also applies to \tfive family models. We conjecture this discrepancy might be due to different annotation schemes adopted by MSMARCO Dev and TREC DL, and will leave this to potential future work.

With the current training recipe, \mamba models achieve competitive performance compared to transformer-based models of similar scales. For example, among models $\approx$ 110M parameters, \texttt{state-spaces/mamba-130m} achieves the highest NDCG@10 on TREC DL19 and DL20 datasets.
\texttt{state-spaces/mamba-370m} also achieves highest NDCG@10 on DL19, and second highest NDCG@10 on DL20 among 5 models we compare with scale of $\approx$ 330M parameters. 
\mamba's strong performance is consistent when trained with LoRA. 
For example, \texttt{state-spaces/mamba-790m} compared to \texttt{T5-large} and \texttt{EleutherAI/pythia-1b}, approaching full parameter fine-tuning of \texttt{state-spaces/mamba-370m}.

\begin{figure*}[t]
\begin{subfigure}{0.65\textwidth}
    \includegraphics[height=5cm]{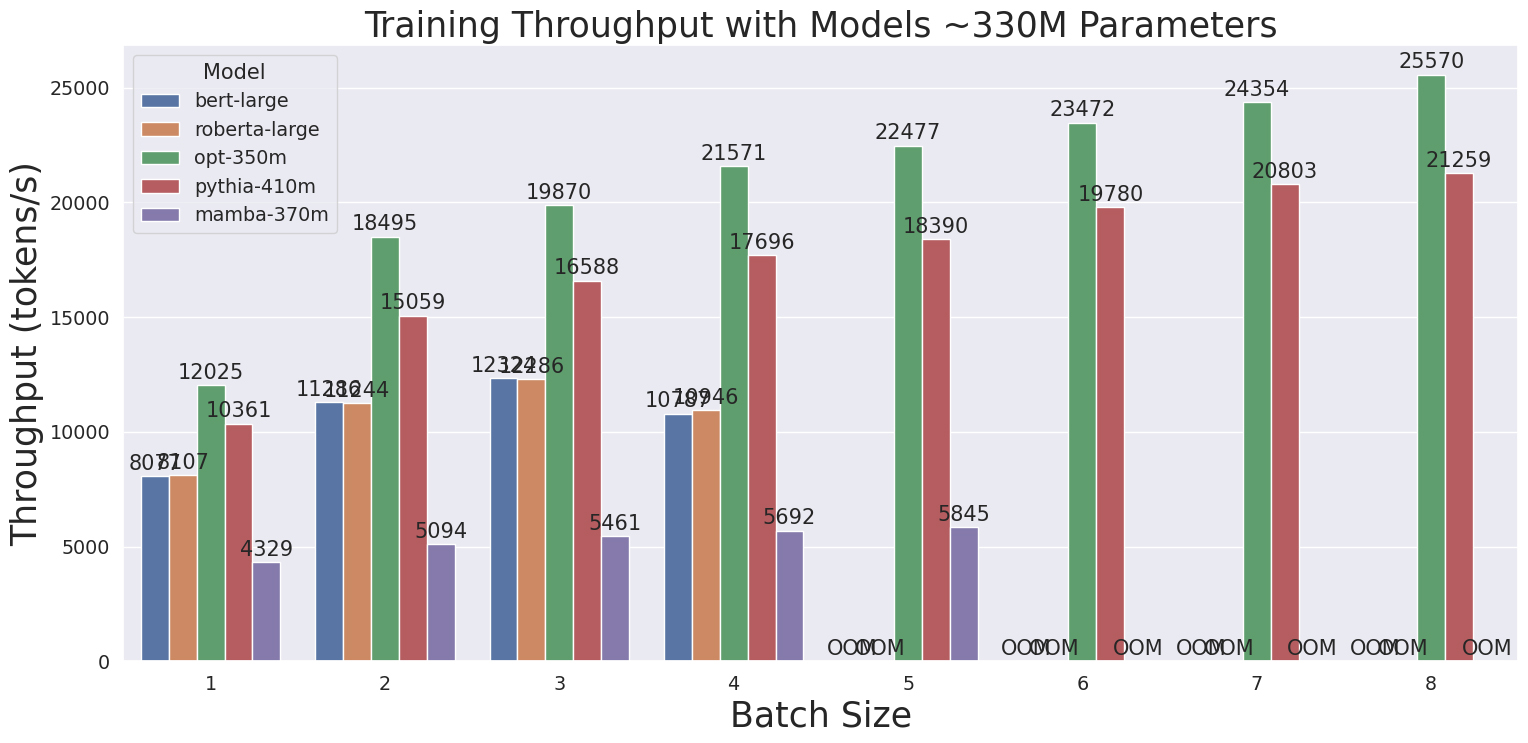}
\end{subfigure}
\hspace{-10pt}
\begin{subfigure}{0.32\textwidth}
    \includegraphics[height=5cm]{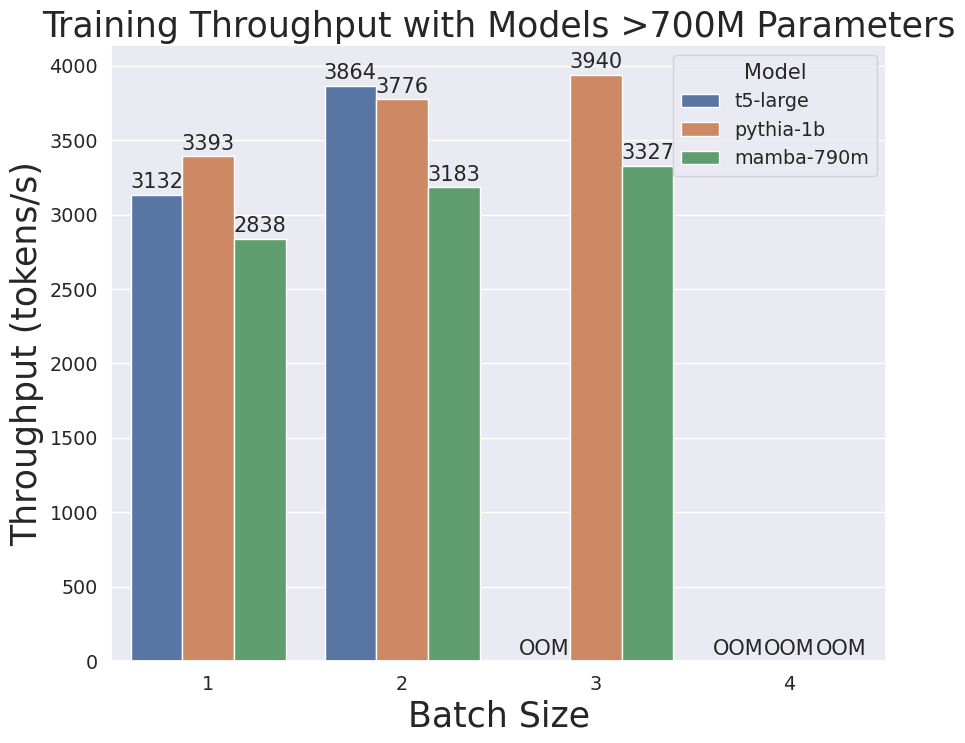}
\end{subfigure}
\caption{
We show the training throughput of models $\approx$330M parameters and models $>$ 700M parameters. Models $>$ 700M parameters are trained with LoRA, with rank=32. We notice \mamba models have lower throughput and higher GPU memory consumption compared to efficient transformer implementations such as flash attention.
}
\label{fig:throughput}
\end{figure*}

\paragraph{Benchmarking Training Throughput for Document Ranking.}
We benchmark the training throughput of different models on a server with one A6000 GPU and report results in Figure~\ref{fig:throughput}.
Overall, models trained with flash attention implementation (\opt, \pythia) achieve the highest training throughput and lowest space complexity. 
We notice that although \mamba can achieve $O(n)$ training complexity in theory, the training throughput with its current implementation is still lower than efficient transformer implementations such as flash attention.
This finding suggests the deficiency of \mamba's current implementation and leaves us with potential future work for improvement.

\section{Related Works}
\label{sec:related}
\paragraph{Transformer-based Language Models} These language models utilize the attention mechanism~\citep{vaswani2017attention} to produce contextualized representations of tokens and demonstrate superior performance and training efficiency compared to traditional recurrent neural network models~\citep{hochreiter1997long,peters-etal-2018-deep}.
Such LMs can be used as feature extractors for downstream tasks or be used for natural language generation (NLG) with a specific language modeling head.
Prior works~\citep{xu2024multi,raffel2020exploring,wang2022language,le-scao-etal-2022-language} have shown that between models of similar sizes, different pre-training objectives can lead to strengths and weaknesses in terms of natural language understanding (NLU) and natural language generation capabilities.

\paragraph{Language Models in IR}
Neural network-based language modeling techniques have gained widespread adoption within the information retrieval (IR) community, including web search snippet generation~\citep{bast2014efficient,xu2023lightweight,xu2023context}, modeling users and products in recommender systems~\citep{xu2021understanding,zeng2021zero}, generating search results explanation~\citep{yu2022towards,xu2023counterfactual} and conversational search~\citep{wang2023depth,wang2023reward,aliannejadi2019asking}.

\paragraph{Document Ranking} The document ranking task aims to capture the relevance between query and document.
Traditional bag-of-words methods such as BM25~\citep{robertson1999okapi} aim to capture the lexical overlapping. 
Neural models such as word2vec~\citep{mikolov2013distributed} are learned from static word co-occurrence patterns from a pre-training corpus and can capture the semantic meanings of the words. 
Transformer-based language models can capture contextualized representations and have been widely applied to modeling query-document relevance. 
\cite{dai2019deeper} propose to model long documents with \bert~\citep{devlin-etal-2019-bert}. To model long documents with \bert's limited 512 context window, they propose several model variants such as FirstP, MaxP and SumP. 
Followup works~\citep{li2023parade,hofstatter2021intra,boytsov2022understanding,ma2023fine} achieve performance improvements by leveraging techniques such as mining hard negatives, chunking long documents and/or sliding windows, and replacing BERT with better, long-context pre-trained language models such as \texttt{Longformer}~\citep{beltagy2020longformer} and \texttt{Llama}~\citep{touvron2023llama,touvron2023llamatwo}. 
For a more comprehensive literature review, refer to \cite{lin2022pretrained} and \cite{xu2025surveymodelarchitecturesinformation}

\section{Conclusion and Limitations}
\label{sec:conclusion}
This study aims to benchmark \mamba models' performance in the document ranking task. We find that \mamba models can achieve competitive performance compared to transformer-based language models, and the effectiveness is consistent with trained with LoRA. 
However, \mamba models have the deficiency of lower training throughput compared to efficient attention implementations, limiting their potential for efficient training and deployment. 
We only explored the contrastive learning training strategy. As recent works demonstrate, knowledge distillation is especially effective for training lightweight rerankers when distilling from a more capable reranker teacher~\citep{baldelli2024twolar,schlatt2025rank,xu2025distillationversuscontrastivelearning}.
Due to the limited bandwidth, we only study reranker models $<$ 1B parameters based on the BM25 retriever.
Our future work should include extending to rerankers of large scales, different document retrievers.
\nocite{xu2025cspladelearnedsparseretrieval}

\bibliographystyle{plainnat}
\bibliography{references}

\end{document}